\documentclass[lettersize,journal]{IEEEtran}
\usepackage{amsmath,amsfonts}
\usepackage{algorithmic}
\usepackage{algorithm}
\usepackage{array}
\usepackage[caption=false,font=normalsize,labelfont=sf,textfont=sf]{subfig}
\usepackage{textcomp}
\usepackage{stfloats}
\usepackage{url}
\usepackage{verbatim}
\usepackage{graphicx}
\usepackage{cite}

\usepackage{amssymb}  
\usepackage{rotating} 
\usepackage{color}    
\begin{document}

\title{\huge Joint Design of Spectrally-Constrained OFDM Sequences and Mismatch Filter With PAPR Constraint}

\author{\IEEEauthorblockN{Jinyang He, Ziyang Cheng, ~\IEEEmembership{Member,~IEEE}  and Zishu He,  ~\IEEEmembership{Member,~IEEE}    \vspace{-2em}
}\\
}



\maketitle

\begin{abstract}
The OFDM  sequences with low correlation sidelobe level (SLCL) is desired in many 5G wireless systems. In this letter,  the OFDM sequences and mismatch filter are jointly designed to achieve the SLCL under the  constraints  of spectra and  peak to average power ratio (PAPR). Specifically, we formulate the optimization problem by maximizing the  peak side lobe ratio (PSLR) of the cross-correlation   between the OFDM  sequences and mismatch filter, subject to several practice constraints.  To solve the nonconvex problem, an efficient alternating optimization (AltOpt) algorithm  is proposed.  Numerical simulations are provided to demonstrate the effectiveness of the proposed algorithms.
\end{abstract}
\vspace{-1em}
\vspace{-1em}
\section{Introduction}
\IEEEPARstart{A}{s} a crucial technique in 4G and 5G wireless communications,  orthogonal frequency division multiplexing (OFDM) signal has been extensively studied \cite{tse2005, molisch2012}. Moreover, due to some excellent performance in radar sensing, 
OFDM   has also been explored for the implementation of beyond 5G(B5G) or 6G, an important feature of which is the integration of sensing functions into communications \cite{Keskin2021, Liu2022}. For example, the derivation for range and velocity measurement utilizing the OFDM sequences is proposed in \cite{Sturm2011}. 
Besides, in \cite{Shi2021}, an OFDM dual-function radar-communication system is achieved by jointly selecting subcarriers and allocating power resources.

Transmit sequences with low correlation sidelobe level (CSLL) play a  crucial part in communication and active sensing applications. For instance, low CSLL can be used in communication systems to facilitate  channel estimation and synchronization performance \cite{Chen2022}, and in radar systems to improve the detection performance of weak targets\cite{Zhou2022}. Therefore, sidelobe suppression has been extensively studied in the literature \cite{Zhou2022,Rabaste2015,Song2015,Zhou2020}. For example,in \cite{Rabaste2015}, the  mismatch filter is designed by minimizing peak sidelobe (PSL) with a given transmit sequence. However, this method suffers from  a loss in the main-lobe gain. Thereby, in \cite{Song2015}, a unimodular sequence is designed by minimizing the  auto-correlation sidelobe level.   Nevertheless, the above two methods only optimize the transmitting sequences or receiving filter separately, and do not fully utilize their degrees of freedom (DoF). In \cite{Zhou2022}, the joint design problem is established by minimizing the weighted sum of ISL and loss-in-processing gain (LPG). This method can obtain a low range sidelobe at a slight cost of LPG.


In addition to the low sidelobe property, the low peak to average power ratio (PAPR) property which can improve the efficiency of the transmitter is also expected \cite{Cheng2018}. In \cite{Tsai2011}, OFDM  sequences with good auto-correlation property and low PAPR under spectral constraints are synthesized. The spectral constraints are frequently necessary to reduce interference in communication systems and prevent the use of bands reserved for navigation or military purposes in radar systems.
In \cite{Tsai2011}, a relaxed method is proposed to find the aperiodic ODFM sequence. However, it will suffer from a performance loss without taking full advantage of the DoF of the receiver.

Motivated by the above facts, in this letter, we investigate the joint design of transmit OFDM sequences and mismatch filter under the  constraints of spectra and PAPR. {The joint design problem is formulated by maximizing the peak side lobe ratio (PSLR) of the  correlation between the OFDM sequences and mismatch filter, subject to the spectral and PAPR constraints. To solve the resultant nonconvex problem, an alternating optimization (AltOpt) algorithm is devised. Specifically, the OFDM sequences is optimized based on the alternating direction method of multipliers (ADMM) framework \cite{Boyd2011} and then the mismatch filter is optimized by generalized eigenvalue decomposition. Representative numerical simulations are provided to illustrate
the performance of the proposed method in terms
of the ISL property.}

\textit{Notation:} $ \mathbf{a} $ and $ \mathbf{A} $ denote a  vector  and a matrix, respectively.    $ (\cdot)^* $,  $ (\cdot)^T $ and $ (\cdot)^H $ separately  stand for the conjugate,   transpose and conjugate transpose operators. $\mathbb{C}^{n}$ and $\mathbb{C}^{N\times N}$ represent the sets of $n$-dimensional complex-valued vectors and $ N \times N $ complex-valued matrices, respectively. The  real part   of a complex-valued number is noted by    $ \Re \left\{  \cdot  \right\} $. 
$  \lVert\cdot\rVert $ denote the  $ l_2 $  norms. Finally,  $\jmath = \sqrt{-1}$ denotes the imaginary unit.

\vspace{-1em}
\section{Problem Formulation}
Consider an OFDM signaling system with $ N $ subcarriers, and $ {\bf s}={{\left[ {{s_1}},\ldots ,{{s_N}} \right]}^{T}} \in {{\mathbb{C}}^{N\times 1}} $ represents the modulated symbol over the $N$ subcarriers. Then, the time-domain sequence can be written as
\begin{equation}
	\begin{aligned}
		\mathbf{x}={{\mathbf{F}}_\mathrm{I}}{\bf s}
	\end{aligned}
	\label{1}
\end{equation}
where ${{\mathbf{F}}_\mathrm{I}\in {{\mathbb{C}}^{N\times N}}}$ is the Inverse-DFT (IDFT) matrix, whose $\left( m,n \right)$-th element is
\begin{equation}
    \begin{aligned}
        {{\mathbf{F}}_\mathrm{I}}\left( m,n \right)=\frac{1}{N}{{e}^{\jmath 2\pi \frac{\left( m-1 \right)\left( n-1 \right)}{N}}}
    \end{aligned}
    \label{2}
\end{equation}

The correlation function between the time-domain sequences $\mathbf{x}$ and mismatch filter $\mathbf{h}={{\left[ {{h}_{1}},{{h}_{2}},\ldots ,{{h}_{N}} \right]}^{T}}\in {{\mathbb{C}}^{N\times 1}}$ is given by
\begin{equation}
    \begin{aligned}
        r_{k}^{xh}=\sum\limits_{n=k+1}^{N}{{{x}_{n}}h_{n-k}^{*}},k=-N+1,\ldots ,N-1
    \end{aligned}
    \label{3}
\end{equation}

Define $\mathbf{r}={{\left[ r_{-N+1}^{xh},\ldots ,r_{-1}^{xh},0,r_{1}^{xh},\ldots ,r_{N-1}^{xh} \right]}^{T}}$ as sidelobe vector, we have $\mathbf{r}={{\mathbf{H}}_{\mathrm{SL}}}\mathbf{x}$, where
\begin{small}
\begin{equation}
    \begin{aligned}
        {{\mathbf{H}}_{\mathrm{SL}}}={{\left[ \begin{matrix}
   {{h}_{N}} & {{h}_{N-1}} & \cdots  & {{h}_{2}} & 0 & 0 & 0 & \cdots  & 0  \\
   0 & {{h}_{N}} & \ddots  & \vdots  & 0 & {{h}_{1}} & \ddots  & \ddots  & \vdots   \\
   0 & 0 & \ddots  & {{h}_{N-1}} & \vdots  & {{h}_{2}} & {{h}_{1}} & \ddots  & 0  \\
   \vdots  & \vdots  & \ddots  & {{h}_{N}} & 0 & \vdots  & \ddots  & \ddots  & 0  \\
   0 & 0 & \cdots  & 0 & 0 & {{h}_{N-1}} & \cdots  & {{h}_{2}} & {{h}_{1}}  \\
\end{matrix} \right]}^{H}}
    \end{aligned}
    \label{4}
\end{equation}
\end{small}

Then, the ISL metric is calculated as 
\begin{equation}
    \begin{aligned}
        \mathrm{ISL}={{\mathbf{r}}^{H}}\mathbf{r}={{\mathbf{x}}^{H}}\mathbf{H}_{\mathrm{SL}}^{H}{{\mathbf{H}}_{\mathrm{SL}}}\mathbf{x}
    \end{aligned}
    \label{5}
\end{equation}

Similarly, the ISL metric can also be written as
\begin{equation}
    \begin{aligned}
        \mathrm{ISL}={{\mathbf{h}}^{H}}\mathbf{X}_{\mathrm{SL}}^{H}{{\mathbf{X}}_{\mathrm{SL}}}\mathbf{h}
    \end{aligned}
    \label{6}
\end{equation}
where
\begin{small}
\begin{equation}
    \begin{aligned}
        {{\mathbf{X}}_{\text{SL}}}={{\left[ \begin{matrix}
   0 & 0 & \cdots  & 0 & 0 & {{x}_{2}} & \cdots  & {{x}_{N-1}} & {{x}_{N}}  \\
   \vdots  & \vdots  & \begin{turn}{90}$\ddots$\end{turn} & {{x}_{1}} & 0 & \vdots  & \begin{turn}{90}$\ddots$\end{turn} & {{x}_{N}} & 0  \\
   0 & 0 & {{x}_{1}} & {{x}_{2}} & \vdots  & {{x}_{N-1}} & \begin{turn}{90}$\ddots$\end{turn} & 0 & 0  \\
   0 & \begin{turn}{90}$\ddots$\end{turn} & \begin{turn}{90}$\ddots$\end{turn} & \vdots  & 0 & {{x}_{N}} & \begin{turn}{90}$\ddots$\end{turn} & \vdots  & \vdots   \\
   {{x}_{1}} & {{x}_{2}} & \cdots  & {{x}_{N-1}} & 0 & 0 & \cdots  & 0 & 0  \\
\end{matrix} \right]}^{H}}
    \end{aligned}
    \label{7}
\end{equation}
\end{small}

Thus, the PSLR is defined as
\begin{equation}
    \mathrm{PSLR} = \frac{{{\mathbf{x}}^{H}}\mathbf{h}{{\mathbf{h}}^{H}}\mathbf{x}}{{{\mathbf{x}}^{H}}\mathbf{H}_{\mathrm{SL}}^{H}{{\mathbf{H}}_{\mathrm{SL}}}\mathbf{x}} = \frac{{{\mathbf{h}}^{H}}\mathbf{x}{{\mathbf{x}}^{H}}\mathbf{h}}{{{\mathbf{h}}^{H}}\mathbf{X}_{\mathrm{SL}}^{H}{{\mathbf{X}}_{\mathrm{SL}}}\mathbf{h}}
\end{equation}

In order to acquire a lower sidelobe and retain the main lobe gain, the PSLR is selected as the optimization criterion. Additionally, {due to some hardware limitations, such as the band-limited pre-filter, some  subcarriers are unavailable. The OFDM sequences are expected to have a low PAPR value to improve  the efficiency of the transmit  power amplifier.  Consequently, the joint design problem of spectrally-constrained OFDM sequences and mismatch filter can be formulated as}
\begin{subequations}
    \begin{align}
        \underset{\mathbf{x},\mathbf{h}}{\mathop{\max }}\,\text{ }  &\frac{{{\mathbf{x}}^{H}}\mathbf{h}{{\mathbf{h}}^{H}}\mathbf{x}}{{{\mathbf{x}}^{H}}\mathbf{H}_{\mathrm{SL}}^{H}{{\mathbf{H}}_{\mathrm{SL}}}\mathbf{x}} \label{8a} \\
  \mathrm{s}\mathrm{.t}\mathrm{.}&\mathbf{x}={{\mathbf{F}}_{\mathrm{I}}}{\bf s} \label{8b} \\ 
 & \left| {{s_n}} \right|=\left\{ \begin{matrix}
   1,n\in \mathcal{N}  \\
   0,n\in \bar{\mathcal{N}}  \\
\end{matrix} \right. \label{8c} \\ 
 & {{\left| {{x}_{n}} \right|}^{2}}\le \rho P,n=1,2,\ldots ,N   \label{8d}
    \end{align}
    \label{8}
\end{subequations}
where equation \eqref{8c} is spectral constraint and \eqref{8d} is PAPR constraint. $\mathcal{N}$ and $\bar{\mathcal{N}}$ denote the available and unavailable subcarriers index. $\rho$ is the parameter to control the level of the PAPR. $P$ is the average power of the OFDM sequence.

Since the above optimization problem involves a nonconvex objective function and a nonconvex spectral constraint. It is NP-hard and challenging to tackle. To this end, in what follows, an alternating optimization (AltOpt) algorithm is devised.

\section{Solution to the optimization problem}
In brief, our proposed AltOpt algorithm for solving the problem \eqref{8} mainly involves the following two procedures:

1) Given the mismatch filter $\mathbf{h}$, the optimal sequence $\mathbf{x}$ is obtained by maximizing the PSLR under the spectral and PAPR constraints.

2) Given the optimal sequences $\mathbf{x}$, the mismatch filter $\mathbf{h}$ is optimized by maximizing the unconstraint PSLR.

\subsection{Optimization of the sequences $\mathbf{x}$}
	
	It can be seen that given the mismatch filter $\mathbf{h}$, the optimal sequences $\mathbf{x}$ can be obtained by solving the following problem
	\begin{subequations}
	   \begin{align}
          \underset{\mathbf{x}}{\mathop{\min }}\,& \text{ }\frac{{{\mathbf{x}}^{H}}\mathbf{H}_{\mathrm{SL}}^{H}{{\mathbf{H}}_\mathrm{SL}}\mathbf{x}}{{{\mathbf{x}}^{H}}\mathbf{h}{{\mathbf{h}}^{H}}\mathbf{x}} \label{9a}\\ 
          \mathrm{s}\mathrm{.t}\mathrm{.}&\text{ }\mathbf{x}={{\mathbf{F}}_\mathrm{I}}{\bf s} \label{9b} \\ 
         & \left| {{s_n}} \right|=\left\{ \begin{matrix}
           1,n\in \mathcal{N}  \\
           0,n\in \bar{\mathcal{N}}  \\
        \end{matrix} \right. \label{9c}\\ 
         & {{\left| {{x}_{n}} \right|}^{2}}\le \rho P,n=1,2,\ldots ,N  \label{9d}
        \end{align}
        \label{9}
	\end{subequations}
    It is seen that this problem involves a nonconvex objective function and a  nonconvex constraint \eqref{9c}. Therefore, it is NP-hard and challenging to tackle. In the following, we devise an efficient algorithm based on the ADMM technique to handle it. To deal with problem \eqref{9}, we first introduce an auxiliary variable $\mathbf{y}=\mathbf{x}$ thus the augmented Lagrangian function (scaled-form) can be written as
    \begin{equation}
        \begin{aligned}
            \mathcal{L}(\mathbf{x},\mathbf{y},\mathbf{u})=\frac{{{\mathbf{x}}^{H}}\mathbf{H}_\mathrm{SL}^{H}{{\mathbf{H}}_\mathrm{SL}}\mathbf{x}}{{{\mathbf{y}}^{H}}\mathbf{h}{{\mathbf{h}}^{H}}\mathbf{y}}+\frac{{{\rho }_{0}}}{2}{{\left\| \mathbf{x}-\mathbf{y}+\mathbf{u} \right\|}^{2}}
        \end{aligned}
        \label{10}
    \end{equation}
    where $\mathbf{u}$ is the dual variable and $\rho_0$ is the penalty parameter.
    
    At the $ (k+1) $th iteration, the ADMM algorithm consists of the following update procedures:
	\begin{subequations}
	   \begin{align}
          &{{\mathbf{x}}^{\left( k+1 \right)}}=\underset{\mathbf{x}\in {{\mathcal{D}}_{x}}}{\mathop{\arg \min }}\,\mathcal{L}\left( \mathbf{x},{{\mathbf{y}}^{\left( k \right)}},{{\mathbf{u}}^{\left( k \right)}} \right) \label{11a}\\ 
         &{{\mathbf{y}}^{\left( k+1 \right)}}=\underset{\mathbf{y}\in {{\mathcal{D}}_{y}}}{\mathop{\arg \min }}\,\mathcal{L}\left( {{\mathbf{x}}^{\left( k+1 \right)}},\mathbf{y},{{\mathbf{u}}^{\left( k \right)}} \right) \label{11b} \\ 
         & {{\mathbf{u}}^{\left( k+1 \right)}}={{\mathbf{u}}^{\left( k \right)}}+{{\mathbf{x}}^{\left( k+1 \right)}}-{{\mathbf{y}}^{\left( k+1 \right)}} \label{11c}
        \end{align}
        \label{11}
	\end{subequations}
    where the sets $ {\cal D}_x $, $ {\cal D}_y $ are, respectively, defined as ${{\mathcal{D}}_{\mathbf{x}}}=\left\{ \mathbf{x}\left| \mathbf{x}={{\mathbf{F}}_\mathrm{I}}{\bf s},\text{ }\left| {{s_n}} \right|=\left\{ \begin{matrix}
   1,n\in \mathcal{N}  \\
   0,n\in \bar{\mathcal{N}}  \\
\end{matrix} \right. \right. \right\}$ and ${{\mathcal{D}}_{\mathbf{y}}}=\left\{ \mathbf{y}\left| {{\left| {{y}_{n}} \right|}^{2}}\le \rho P,n=1,2,\ldots ,N \right. \right\}$.

    In the sequel, the solutions to the alternating minimization problems from \eqref{11a} to \eqref{11c} are presented.
    
    \textit{1) Update $\mathbf{x}$:} Given $\left\{ {{\mathbf{y}}^{\left( k \right)}},{{\mathbf{u}}^{\left( k \right)}} \right\}$ at the $ k $th iteration, $\mathbf{x}^{(k+1)}$ is attained by solving
    \begin{equation}
        \begin{aligned}
  & \underset{\mathbf{x}}{\mathop{\min }}\,\text{ }\frac{{{\mathbf{x}}^{H}}\mathbf{H}_\mathrm{SL}^{H}{{\mathbf{H}}_\mathrm{SL}}\mathbf{x}}{{{\mathbf{y}}^{\left( k \right)}}^{H}\mathbf{h}{{\mathbf{h}}^{H}}{{\mathbf{y}}^{\left( k \right)}}}+\frac{{{\rho }_{0}}}{2}{{\left\| \mathbf{x}-{{\mathbf{y}}^{\left( k \right)}}+{{\mathbf{u}}^{\left( k \right)}} \right\|}^{2}} \\ 
 & \text{ }s.t.\text{  }\mathbf{x}={{\mathbf{F}}_\mathrm{I}}{\bf s},\text{ }\left| {{s_n}} \right|=\left\{ \begin{matrix}
   1,n\in \mathcal{N}  \\
   0,n\in \bar{\mathcal{N}}  \\
\end{matrix} \right. 
        \end{aligned}
        \label{12}
    \end{equation}
    which is nonconvex due to the nonconvex constraint. Nevertheless, the problem can be solved efficiently by the Block Coordinate Descent(BCD) method \cite{Nesterov2012, Hong2016}. More concretely, the optimization of \eqref{12} with respect to $ {s_n} $ is given by
    \begin{equation}
        \begin{aligned}
   \underset{{{s_n}}}{\mathop{\min }}\,&~\kappa \left( {s_n} \right)=\Re \left\{ {s_n}d \right\} \\ 
  {\rm s.t.} &~{s_n} = \left\{ \begin{array}{l}
{e^{\jmath {\phi _n}}},{\phi _n} \in \left[ {0,2\pi } \right],n \in {\cal N}\\
0,\qquad \quad \qquad \qquad n \in \overline {\cal N} 
\end{array} \right.
        \end{aligned}
        \label{13}
    \end{equation}
    where $d$ is defined as
    \begin{equation}
        \begin{aligned}
            d=&\frac{2{{\mathbf{\bar{s}}}_{n}^{H}}\mathbf{F}_
        \mathrm{I}^{H}\mathbf{H}_\mathrm{SL}^{H}{{\mathbf{H}}_\mathrm{SL}}{{\mathbf{F}}_\mathrm{I}}\mathbf{\tilde{e}}_n}{{{\mathbf{y}}^{\left( k \right)}}^{H}\mathbf{h}{{\mathbf{h}}^{H}}{{\mathbf{y}}^{\left( k \right)}}}
            +{{\rho }_{0}}{{\left( {{\mathbf{F}}_\mathrm{I}}\mathbf{\bar{s}}_{n}-{{\mathbf{y}}^{\left( k \right)}}+{{\mathbf{u}}^{\left( k \right)}} \right)}^{H}}{{\mathbf{F}}_\mathrm{I}}\mathbf{\tilde{e}}_n
        \end{aligned}
        \label{14}
    \end{equation}
     with  $ {\mathbf{\bar{s}}}_{n}  $ being the vector  $ {\bf{s}} $ whose $n$-th entry is zeroed and $ {{\bf{\tilde e}}_n} $ being an $N\times 1$ vector, whose  $n$-th entry is 1, and 0 for otherwise.
     
     Thus, it is easy to obtain the optimal solution to the problem \eqref{14} as
     \begin{equation}
         \begin{aligned}
              {{s_n}}=\left\{ \begin{matrix}
   -{{e}^{-\jmath\arg \left( d \right)}},&n\in \mathcal{N}  \\
   0,&n\in \bar{\mathcal{N}}  \\
\end{matrix} \right.
         \end{aligned}
         \label{15}
     \end{equation}
     
     \textit{2) Update $\mathbf{y}$:} For given $\left\{ {{\mathbf{x}}^{\left( k+1 \right)}},{{\mathbf{u}}^{\left( k \right)}} \right\}$, $ {\bf{y}} $ can be obtained by solving the following minimization problem
     \begin{equation}
         \begin{aligned}
           \underset{\mathbf{y}}{\mathop{\min }}\,&\frac{{{\mathbf{x}}^{\left( k+1 \right)}}^{H}\mathbf{H}_\mathrm{SL}^{H}{{\mathbf{H}}_\mathrm{SL}}{{\mathbf{x}}^{\left( k+1 \right)}}}{{{\mathbf{y}}^{H}}\mathbf{h}{{\mathbf{h}}^{H}}\mathbf{y}}+\frac{{{\rho }_{0}}}{2}{{\left\| {{\mathbf{x}}^{\left( k+1 \right)}}-\mathbf{y}+{{\mathbf{u}}^{\left( k \right)}} \right\|}^{2}} \\ 
         \mathrm{s}\mathrm{.t}\mathrm{.}&{{\left| {{y}_{n}} \right|}^{2}}\le \rho P,n=1,2,\ldots ,N \\ 
         \end{aligned}
         \label{16}
     \end{equation}
     Since the problem \eqref{16} is nonconvex, it is challenging to obtain a closed-form solution. However, to reduce the complexity, the closed-form solution is still desired. To this end, we propose to derive a suboptimal solution by a two-step strategy. Specifically, we first obtain the optimal solution to the unconstrained problem \cite{Deng2022} and then project it onto the feasible set. The unconstrained version of the problem \eqref{16} is given by
     \begin{equation}
         \begin{aligned}
              \underset{\mathbf{y}}{\mathop{\min }}\,\frac{{{\mathbf{x}}^{\left( k+1 \right)}}^{H}\mathbf{H}_\mathrm{SL}^{H}{{\mathbf{H}}_\mathrm{SL}}{{\mathbf{x}}^{\left( k+1 \right)}}}{{{\mathbf{y}}^{H}}\mathbf{h}{{\mathbf{h}}^{H}}\mathbf{y}}+\frac{{{\rho }_{0}}}{2}{{\left\| {{\mathbf{x}}^{\left( k+1 \right)}}-\mathbf{y}+{{\mathbf{u}}^{\left( k \right)}} \right\|}^{2}}
         \end{aligned}
         \label{17}
     \end{equation}
     To solve the problem \eqref{17}, we first perform the eigen-decomposition of $\mathbf{h}{{\mathbf{h}}^{H}}$ as $\mathbf{h}{{\mathbf{h}}^{H}}=\mathbf{U\Lambda }{{\mathbf{U}}^{H}}$, where $\mathbf{U}$ is the eigenmatrix and $\mathbf{\Lambda }=\mathrm{diag}\left\{ {{\lambda }_{1}},{{\lambda }_{2}},\ldots ,{{\lambda }_{N}} \right\}$ is the eigen-value matrix with ${{\lambda }_{1}}\ge {{\lambda }_{2}}\ge \ldots \ge {{\lambda }_{N}}$.
     
     Defining $\mathbf{\tilde{y}}={{\mathbf{U}}^{H}}\mathbf{y}$ and $\mathbf{q}={{\mathbf{U}}^{H}}\left( {{\mathbf{x}}^{\left( k+1 \right)}}+{{\mathbf{u}}^{\left( k \right)}} \right)$, the problem \eqref{17} is equivalent to 
     \begin{equation}
         \begin{aligned}
            \underset{{\mathbf{\tilde{y}}}}{\mathop{\min }}\,\frac{{{\mathbf{x}}^{\left( k+1 \right)}}^{H}\mathbf{H}_\mathrm{SL}^{H}{{\mathbf{H}}_\mathrm{SL}}{{\mathbf{x}}^{\left( k+1 \right)}}}{{{{\mathbf{\tilde{y}}}}^{H}}\mathbf{\Lambda \tilde{y}}}+\frac{{{\rho }_{0}}}{2}{{\left\| \mathbf{\tilde{y}}-\mathbf{q} \right\|}^{2}}
         \end{aligned}
        \label{18}
     \end{equation}
    
    Since $\mathbf{h}{{\mathbf{h}}^{H}}$ is rank-one as defined below \eqref{17}, the corresponding eigenvalues are as follows
    \begin{equation}
        \begin{aligned}
            {{\lambda }_{1}}>0,{{\lambda }_{k}}=0,k=2,3,\ldots ,N
        \end{aligned}
        \label{19}
    \end{equation}
    Thus, for $k=2,3,\ldots ,N$, the optimal ${{\tilde{y}}_{k}}={{q}_{k}}$, where ${\tilde{y}}_{k}$ and ${q}_{k}$ are the $k$-th elements of $\mathbf{\tilde{y}}$ and $\mathbf{q}$.
    
    To obtain the optimal ${{\tilde{y}}_{1}}$, we turn \eqref{18} to a real-value problem, by letting
    \begin{equation}
        \begin{aligned}
            {{\tilde{y}}_{1r}} = {\Re \left( {{\tilde{y}}_{1}} \right)}, {{\tilde{y}}_{1i}} = {\Im \left( {{\tilde{y}}_{1}} \right)},\\
            {{{q}}_{1r}} = {\Re \left( {{{q}}_{1}} \right)}, {{{q}}_{1i}} = {\Im \left( {{{q}}_{1}} \right)}.
        \end{aligned}
        \label{20}
    \end{equation}
    
    We can rewritten \eqref{18} as real-value problem,
    \begin{equation}
        \begin{aligned}
            \underset{{{{\tilde{y}}}_{1r}},{{{\tilde{y}}}_{1i}}}{\mathop{\min }}\,&\frac{{{\mathbf{x}}^{\left( k+1 \right)}}^{H}\mathbf{H}_\mathrm{SL}^{H}{{\mathbf{H}}_\mathrm{SL}}{{\mathbf{x}}^{\left( k+1 \right)}}}{{{\lambda }_{1}}\left( \tilde{y}_{1r}^{2}+\tilde{y}_{1i}^{2} \right)}\\
            &+\frac{{{\rho }_{0}}}{2}\left[ {{\left( {{{\tilde{y}}}_{1r}}-{{q}_{1r}} \right)}^{2}}+{{\left( {{{\tilde{y}}}_{1i}}-{{q}_{1i}} \right)}^{2}} \right]
        \end{aligned}
        \label{21}
    \end{equation}
    
    Let
    \begin{equation}
        \begin{aligned}
            \tilde{f}\left( {{{\tilde{y}}}_{1r}},{{{\tilde{y}}}_{1i}} \right)\triangleq \frac{p}{\left( \tilde{y}_{1r}^{2}+\tilde{y}_{1i}^{2} \right)}+{{\left( {{{\tilde{y}}}_{1r}}-{{q}_{1r}} \right)}^{2}}+{{\left( {{{\tilde{y}}}_{1i}}-{{q}_{1i}} \right)}^{2}}
        \end{aligned}
        \label{22}
    \end{equation}
    then we obtain
    the first-order optimality condition of $\tilde{f}\left( {{{\tilde{y}}}_{1r}},{{{\tilde{y}}}_{1i}} \right)$ as 
    \begin{subequations}
    \begin{align}
  & \frac{\partial \tilde{f}\left( {{{\tilde{y}}}_{1r}},{{{\tilde{y}}}_{1i}} \right)}{\partial {{{\tilde{y}}}_{1r}}}\triangleq \frac{-2p{{{\tilde{y}}}_{1r}}}{{{\left( \tilde{y}_{1r}^{2}+\tilde{y}_{1i}^{2} \right)}^{2}}}+2\left( {{{\tilde{y}}}_{1r}}-{{q}_{1r}} \right)=0 \label{23a} \\ 
 & \frac{\partial \tilde{f}\left( {{{\tilde{y}}}_{1r}},{{{\tilde{y}}}_{1i}} \right)}{\partial {{{\tilde{y}}}_{1i}}}\triangleq \frac{-2p{{{\tilde{y}}}_{1i}}}{{{\left( \tilde{y}_{1r}^{2}+\tilde{y}_{1i}^{2} \right)}^{2}}}+2\left( {{{\tilde{y}}}_{1i}}-{{q}_{1i}} \right)=0 \label{23b}
\end{align}
\label{23}
    \end{subequations}
    After some mathematical operations, we acquire ${{q}_{1i}}{{\tilde{y}}_{1r}}={{q}_{1r}}{{\tilde{y}}_{1i}}$. Then the following steps allows to obtain the optimal ${\tilde{y}}_{1r}$ and ${\tilde{y}}_{1i}$.
    
    \begin{enumerate}
    \item{If ${{q}_{1i}}={{q}_{1r}}=0$, then $\tilde{y}_{1r}^{2}+\tilde{y}_{1i}^{2}=\sqrt{p}$ according to \eqref{23}. Without prior information, the optimal solution can be randomly generated which satisfies $\tilde{y}_{1r}^{2}+\tilde{y}_{1i}^{2}=\sqrt{p}$.}
    \item{If ${{q}_{1r}}\ne 0,{{q}_{1i}}=0$, then ${{\tilde{y}}_{1i}}=0$. Substituting ${{\tilde{y}}_{1i}}=0$ into \eqref{23}, the optimal ${{\tilde{y}}_{1r}}$ can be obtained by solving $\tilde{y}_{1r}^{4}-{{q}_{1r}}\tilde{y}_{1r}^{3}-p=0$ whose roots are calculated by the formula of computing roots \cite{Shmakov2011} or software like Matlab. Note that there may be more than one real root of the equation, and select the one
    that minimizes $\tilde{f}\left( {{{\tilde{y}}}_{1r}},{{{\tilde{y}}}_{1i}} \right)$.}
    \item{If ${{q}_{1r}}=0,{{q}_{1i}}\ne 0$, which is analogous to case 2). We have ${{\tilde{y}}_{1r}}=0$ and need to to solve $\tilde{y}_{1i}^{4}-{{q}_{1i}}\tilde{y}_{1i}^{3}-p=0$.}
    \item{If ${{q}_{1r}}\ne 0,{{q}_{1i}}\ne 0$, which is the most typical case. Acoording to \eqref{23}, the optimal ${{\tilde{y}}_{1r}}$ can be acquired by solving
    \begin{equation}
        \begin{aligned}
            \tilde{y}_{1r}^{4}-{{q}_{1r}}\tilde{y}_{1r}^{3}-\frac{pq_{1r}^{4}}{{{\left( q_{1r}^{2}+q_{1i}^{2} \right)}^{2}}}=0,
        \end{aligned}
        \label{24}
    \end{equation}
    whose  optimal solution  is ${{\tilde{y}}_{1i}}={{q}_{1i}}{{\tilde{y}}_{1r}}/{{q}_{1r}}$.
    }
    \end{enumerate}
    
    At last, we obtain the optimal solution $\mathbf{y}=\mathbf{U\tilde{y}}$ to the unconstrained problem \eqref{17}. The closed-form suboptimal solution to the problem \eqref{16} is given by \cite{Cheng2018Com, Cheng2019}
    \begin{equation}
        \begin{aligned}
            y_{n}^{\left( k+1 \right)}=\left\{ \begin{matrix}
   \frac{\sqrt{\rho P}{{y}_{n}}}{\left| {{y}_{n}} \right|},&{{\left| {{y}_{n}} \right|}^{2}}>\rho P  \\
   {{y}_{n}},&\mathrm{otherwise}  
\end{matrix} \right.
        \end{aligned}
        \label{25}
    \end{equation}
    
\subsection{Optimization of the mismatch filter $\mathbf{h}$}
According to \eqref{8}, it is known for a fixed sequences $\mathbf{x}$, the maximization of \eqref{8a} with respect to $\mathbf{h}$ can be written as
\begin{equation}
    \begin{aligned}
    \underset{\mathbf{h}}{\mathop{\max }}\,\text{ }\frac{{{\mathbf{h}}^{H}}\mathbf{x}{{\mathbf{x}}^{H}}\mathbf{h}}{{{\mathbf{h}}^{H}}\mathbf{X}_\mathrm{SL}^{H}{{\mathbf{X}}_\mathrm{SL}}\mathbf{h}}
    \end{aligned}
    \label{26}
\end{equation}
where $\mathbf{X}_\mathrm{SL}$ is defined in \eqref{7}. It is seen that the problem in \eqref{26} is an unconstrained maximization problem, whose optimal solution is the maximum generalized eigenvalue of $\mathbf{x}{{\mathbf{x}}^{H}}$ and $\mathbf{X}_\mathrm{SL}^{H}{{\mathbf{X}}_\mathrm{SL}}$.

Based on the above analysis, the overall algorithm for solving problem \eqref{8} is summarized in Algorithm \ref{alg1}.
\begin{algorithm}[H]
\caption{AltOpt Algorithm for Solving \eqref{8}}\label{alg:alg1}
\begin{algorithmic}
\STATE 
\STATE {\bf{Input:}} Initialize $\mathbf{x}$, $\mathbf{h}$, auxiliary variable $\mathbf{y}$, dual variable $\mathbf{u}$, and the max iteration number $M$.
\STATE {\bf{for}} $m=1,\cdots,M$
\STATE \hspace{0.5cm} Set $k=1$
\STATE \hspace{0.5cm} {\bf{while}}
\STATE \hspace{1cm} Update $\mathbf{x}^{(k+1)}$ by the BCD method.
\STATE \hspace{1cm} Update $\mathbf{y}^{(k+1)}$ by \eqref{25}.
\STATE \hspace{1cm} Update $\mathbf{u}^{(k+1)}$ by ${{\mathbf{u}}^{\left( k+1 \right)}}={{\mathbf{u}}^{\left( k \right)}}+{{\mathbf{x}}^{\left( k+1 \right)}}-{{\mathbf{y}}^{\left( k+1 \right)}}$ \\
\STATE \hspace{1cm} $k=k+1$.
\STATE \hspace{0.5cm} {\bf{end while until convergence}}
\STATE \hspace{0.5cm} Update $\mathbf{h}^{(m+1)}$ by finding the maximum generalized\\ \hspace{0.5cm}  eigenvalue of $\mathbf{x}{{\mathbf{x}}^{H}}$ and $\mathbf{X}_\mathrm{SL}^{H}{{\mathbf{X}}_\mathrm{SL}}$.
\STATE {\bf{end for}} $m = M$.
\STATE {\bf{Output:}} transmit sequences $\mathbf{x}$ and mismatch filter $\mathbf{h}$.
\end{algorithmic}
\label{alg1}
\end{algorithm}

Note that it is important to evaluate the computational complexity of the Algorithm \ref{alg1}.
For each optimization of the sequences $ \mathbf{x} $, the main computational complexity of the ADMM method is caused by applying the BCD method to solving \eqref{14} with a complexity of  $ {\mathcal O}(I_1 N^3) $, where $ I_1  $ is the number of iterations required in the BCD method. Then, the optimization of  $ \mathbf{h} $ based on eigen value decomposition requires a complexity of $ {\mathcal O}(N^3) $. To summarize, the overall complexity of the AltOpt algorithm is $ {\mathcal O}(M I_1 I_2 N^3 ) $, where $ I_2 $ is the number of ADMM iterations.

\section{Numerical Simulations}
In this section, we assess the performance of the proposed joint design algorithm 
for spectrally-constrained OFDM signals. Suppose there are 512 subcarriers and 96 of them are nulled at the guard bands. The set of indexes of these null subcarriers is $\bar{\mathcal{N}} = \left\{ 209,210,\ldots ,304 \right\}$. The primal variable ${\bf s}^{(1)}$ is initialized with random-phase sequence. The initial dual variable is set to $\mathbf{z}^{(1)} = \mathbf{0}$. The max iteration number $M = 2000$ and the penalty parameter is $\rho_0 = 10$.

\begin{figure}[!t]
    \centering
    \includegraphics[width=0.8\linewidth]{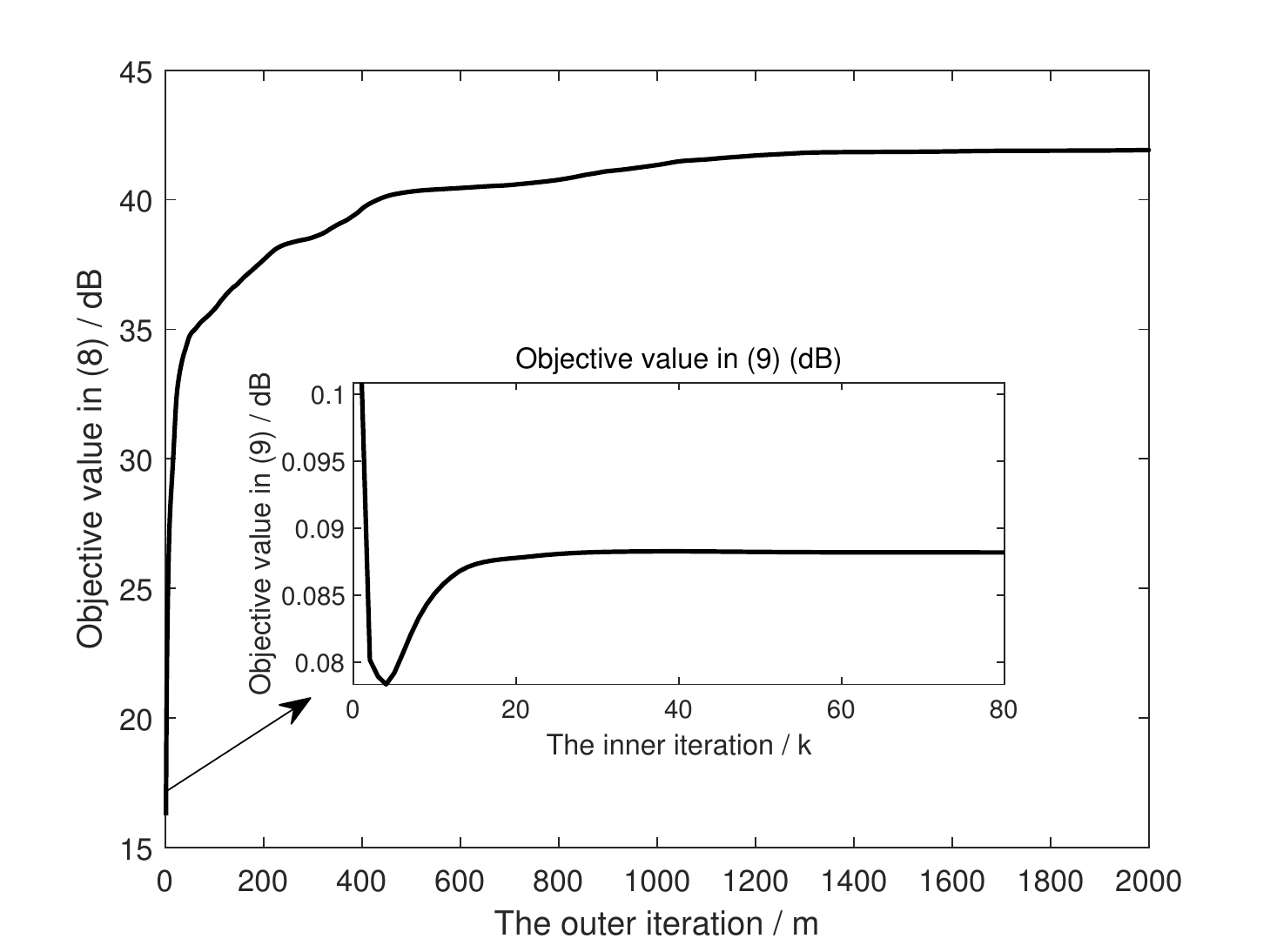}
    \caption{ The objective function value versus iteration. }
   \label{fig:pic1}\vspace{-1em}
\end{figure}

Fig. \ref{fig:pic1} shows the convergence performance of the proposed AltOpt algorithm with  the PAPR level being $\rho=1.25$. It is found that the resulting
objective function in \eqref{8} can converge as the
outer iteration increasing. In addition,
for an instance of the AltOpt framework, we also plot the
objective value in \eqref{9} versus the inner iteration
number by using the ADMM algorithm. 
The result shows that the objective value obtained by the ADMM is able to converge to a sub-optimal value.

\begin{figure}[!t]
    \centering
    \includegraphics[width=0.8\linewidth]{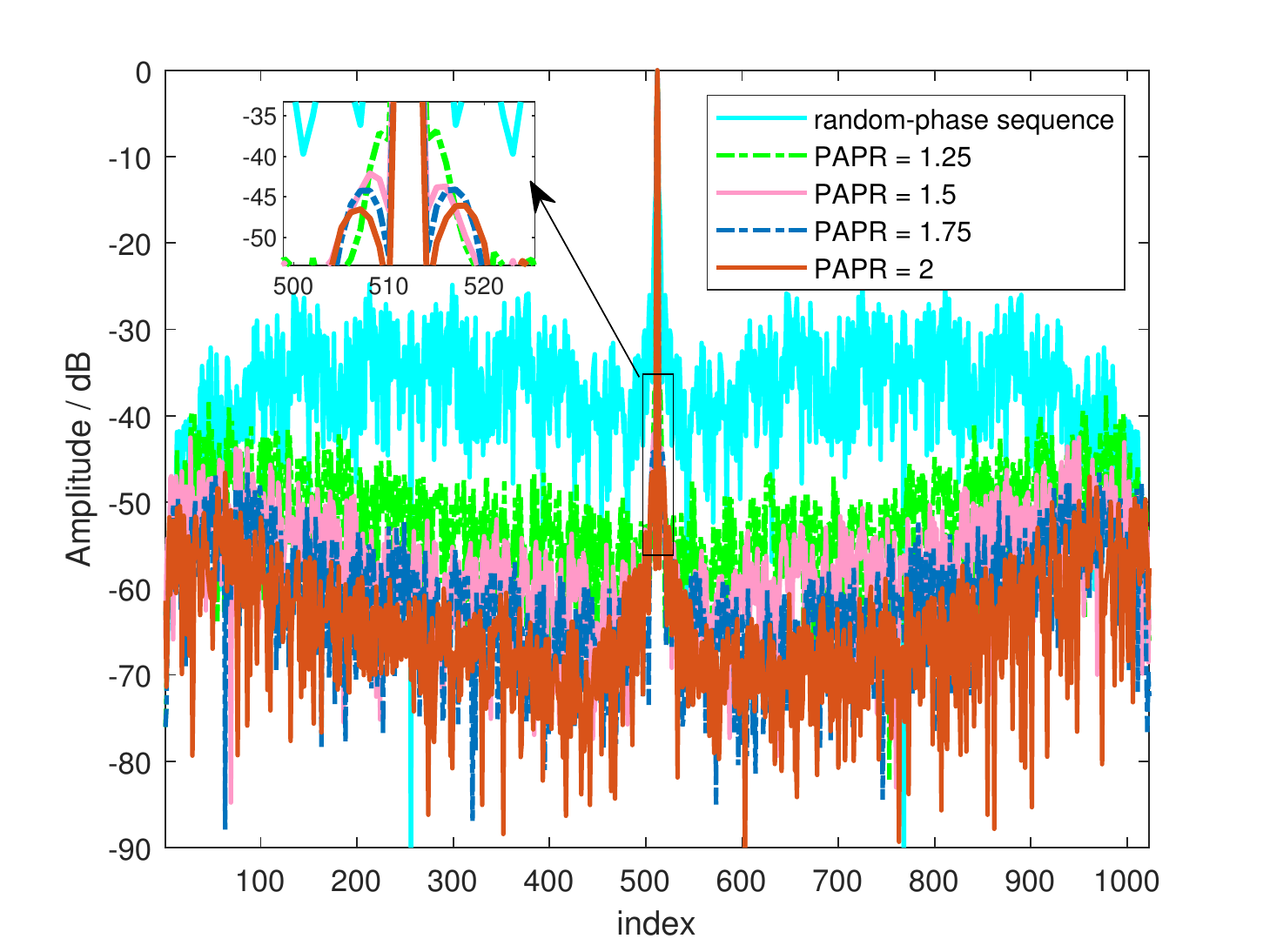}
    \caption{ Cross-correlation levels between the transmitted sequences and the mismatched filter for different PAPR levels.}
   \label{fig:pic2}\vspace{-1em}
\end{figure}

Fig. \ref{fig:pic2} displays the correlation values between the designed sequences and mismatch filter for different PAPR levels. As expected,  the sidelobe level decreases with the PAPR level increasing, since the higher the PAPR level,  the more degrees of freedom (DoFs) can be used to maximize the objective. Moreover, we also observe the designed sequences outperform the random-phase sequences in terms of the ISL property.

\begin{figure}[!t]
    \centering
    \includegraphics[width=0.8\linewidth]{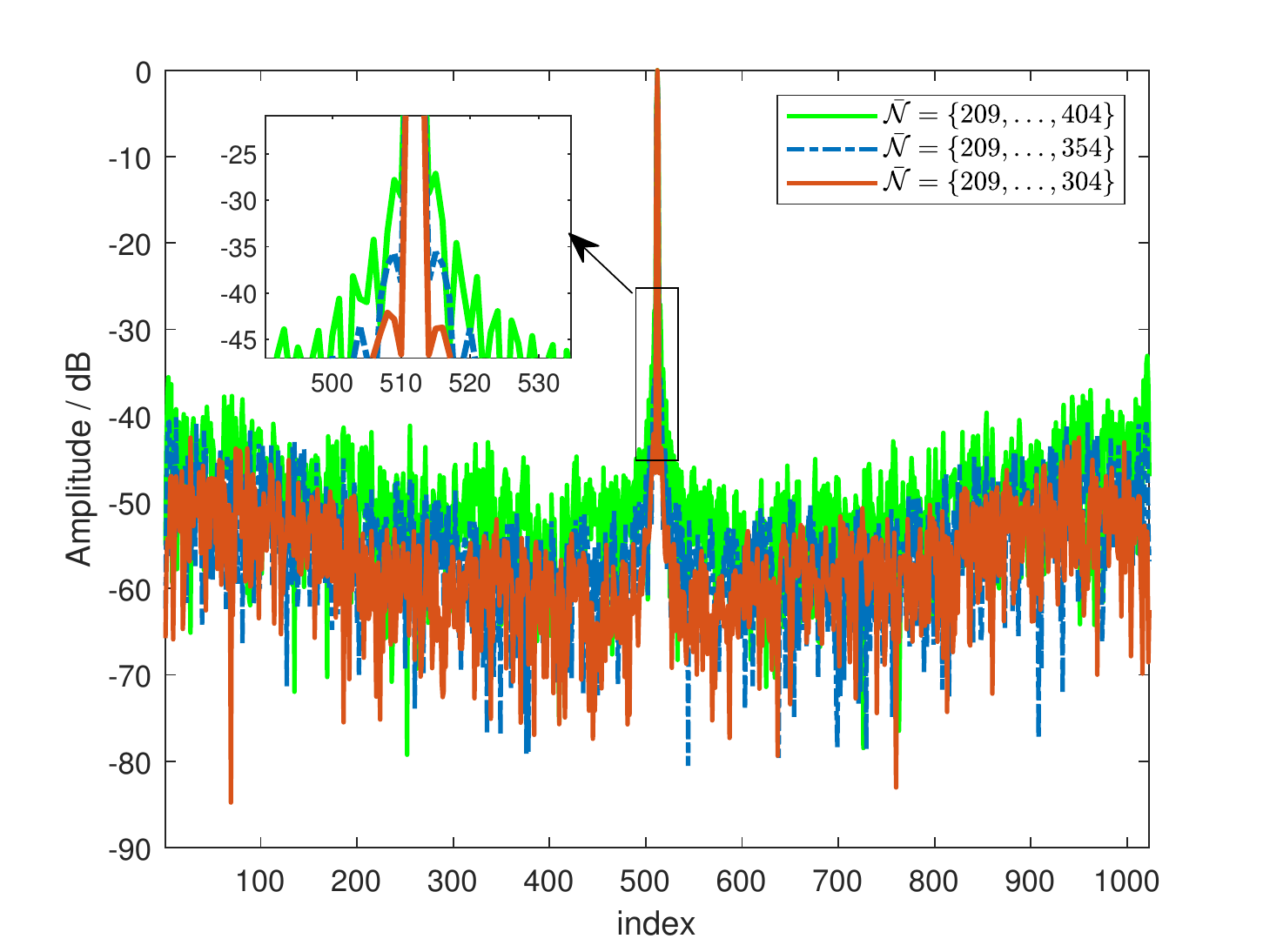}
    \caption{ Cross-correlation levels between the transmitted sequences and the mismatched filter for various numbers of unavailable subcarriers. }
   \label{fig:pic3}\vspace{-1em}
\end{figure}

Finally, the correlation results for various numbers of unavailable subcarriers are shown in Fig. \ref{fig:pic3}.  It is seen that  when the number of  available subcarriers  increases, better ISL properties can be achieved, which is consistent with our expectation that the more the number of available subcarriers, the more DOFs can be used to achieve better ISL properties.

\section{Conclusion}
In this letter, we have addressed the problem of the joint design of transmit sequences and mismatch filter for spectrally-constrained OFDM signals with the PAPR requirement. To tackle the non-convex optimization problem, we have devised an AltOpt algorithm based on the ADMM technique. The performance of the proposed method in terms of convergence and correlation results has been evaluated through numerical simulations. The results show that the proposed method is able to achieve an excellent trade-off among the ISL property, PAPR level and spectral limitation.

 \bibliographystyle{IEEEtran}
\bibliography{IEEEabrv,stan_ref}

\end{document}